\documentclass[
aip,
jcp,
preprint,
unsortedaddress,
amsmath,amssymb,amsfonts,
citeautoscript
]{revtex4-1}

\usepackage{graphicx}  
\usepackage{dcolumn}   
\usepackage{bm}        
\usepackage{amssymb}   
\usepackage{amsfonts}
\usepackage{amsmath}

\begin{document}

\title{Revision of the classical nucleation theory for supersaturated solutions}

\author{Alexander~Borysenko}
\email{borisenko@kipt.kharkov.ua}

\affiliation{                    
  National Science Center "Kharkiv Institute of Physics and Technology", Akademichna Street 1, 61108 Kharkiv, Ukraine
}


\begin{abstract}
During the processes of nucleation and growth of a precipitate cluster from a supersaturated solution, the diffusion flux between the cluster and the solution changes the solute concentration near the cluster-solution interface from its average bulk value. This feature affects the rates of attachment and detachment of solute atoms at the interface and, therefore, alters the entire nucleation kinetics. Unless quite obvious, this effect has been ignored in the classical nucleation theory. To illustrate the results of this new approach, for the case of homogeneous nucleation, we calculate the total solubility (including the contribution from heterophase fluctuations) and the nucleation rate as functions of two parameters of the model and compare these results to the classical ones. One can conclude that discrepancies with the classical nucleation theory are great in the diffusion-limited regime, when the bulk diffusion mobility of solute atoms is small compared to the interfacial one, while in the opposite interface-limited case they vanish.

\end{abstract}

\maketitle

Normally, the process of precipitation from a solution within the binodal-spinodal gap occurs along the 
nucleation-growth-coarsening sequence. Contrary to polymorphic phase transitions (e.g. freezing and condensation), the rate of which is controlled only by the interfacial kinetics, the rate of precipitation from solutions is dependent also on the bulk diffusion kinetics. The diffusion flux between the precipitating cluster and the ambient solution creates a non-uniform solute concentration profile around the cluster. If a cluster is subcritical, the diffusion flux is directed outside of it and the local solute concentration near the interface exceeds the average one. In the case of a supercritical cluster the diffusion flux is directed inside it and the local solute concentration near the interface is less than the average one. This perturbation of the solute concentration profile around the clusters has been observed both in experiments~\cite{Zhang} and simulations~\cite{Kelton_et_al}. Obviously, this fact has to be taken into account when one calculates the nucleation barrier. Nevertheless, this perturbation is often neglected and the solute concentration outside of the clusters is set equal to the average one (see e.g., Ref.~\onlinecite{Slezov} and references therein). As a counterexample, one can refer to the coupled-flux nucleation model elaborated by Kelton~\cite{Kelton}, which considers the intermediate nearest-neighbor shell between the cluster and the solution, which exchanges solute atoms with both adjacent phases. But again, the solute concentration outside of the shell is set equal to the average one, giving a somewhat artificial "staircase" picture of the solute concentration profile. In this paper, working in the framework of the classical nucleation theory (CNT), we consider the kinetics of solute precipitation, taking into account the local perturbation of the solute concentration profile around the clusters and compare our results to the classical ones.

The rate of precipitation, being the first-order phase transition, is conventionally described by the value of the flux of clusters in the dimension space:
\begin{eqnarray} \label{J}
J_{n, n+1}= w^{(+)}_{n, n+1} g \left( n, t \right) - w^{(-)}_{n+1, n} g \left( n+1, t \right),
\end{eqnarray}
where $g \left( n, t \right)$ is a time-dependent distribution function (a concentration of the clusters consisting of $n$ atoms) and $w^{(+)}_{n, n+1}$ and 
$w^{(-)}_{n+1, n}$ are respectively the rates of attachment 
and detachment of solute atoms at the cluster-solution 
interface. The special case when the flux~(\ref{J}) is zero for any $n$: 
\begin{eqnarray} \label{J_0}
J_{n, n+1}=0,~\forall n
\end{eqnarray}
corresponds to the state of detailed balance, when the nucleating phase is in equilibrium with the solution. 
The central postulate of the classical nucleation theory says that the equilibrium (corresponding to the state of detailed balance~(\ref{J_0})) distribution function $g_{\rm{eq}}\left(n\right)$  has the Boltzmann form:
\begin{eqnarray} \label{Boltzmann_distr}
g_{\rm{eq}}\left(n\right)=C\exp\left[-\Delta G\left(n\right)/k_{\rm{B}}T\right],
\end{eqnarray}
where $C$ is a normalization constant, $k_{\rm{B}}$ is the Boltzmann's constant, $T$ is temperature and $\Delta G\left(n\right)$ is a change of the thermodynamic potential of the system as a result of clusterization. Supposed that the cluster is a sphere with a radius of $r_{\rm{n}}=\sqrt[3]{3n\omega_{\rm{n}}/4\pi}$ ($\omega_{\rm{n}}$ being the volume per atom of the cluster), in the capillary approximation the change of the thermodynamic potential is as follows:
\begin{eqnarray} \label{R}
\Delta G\left(n\right)=\int_{0}^{n} \, \left( \mu_{\rm{n}} - \mu_{\rm{s}} \right) \, \mathrm{d} n' + 4\pi r_{\rm{n}}^2 \sigma.
\end{eqnarray}
Here $\mu_{\rm{n}}$ is the chemical potential of a clusterized atom, $\sigma$ is a coefficient of tension at the cluster interface and $\mu_{\rm{s}}$ is the chemical potential of a dissolved atom  near the interface: 
\begin{eqnarray} \label{mu_s}
\mu_{\rm{s}}=\psi + k_{\rm{B}}T \ln c\left(r_{\rm{n}} \right),
\end{eqnarray}
where $\psi$ is a function of pressure and temperature and $c\left(r_{\rm{n}} \right)$ is a molar fraction (concentration) of dissolved atoms near the interface. 
Therefore, in the state of detailed balance~(\ref{J_0}), with the distribution function given by Eq.~(\ref{Boltzmann_distr}) and the work of nucleation given by Eq.~(\ref{R}), from Eq.~(\ref{J}) one can derive a ratio of the attachment and detachments rates as follows:
\begin{align} \label{ratio}
\frac{w^{(+)}_{n, n+1}}{w^{(-)}_{n+1, n}}=&\exp\left[-\frac{\Delta G\left(n+1\right)-\Delta G\left(n\right)}{k_{\rm{B}}T}\right] \nonumber \\
\underset{n \gg 1}{=} &\exp\left[-\frac{\mathrm{d} \Delta G\left(n\right)}{\mathrm{d} n \cdot k_{\rm{B}}T}\right]=\frac{c\left(r_{\rm{n}} \right)}{c_{\rm{eq}}\left(r_{\rm{n}} \right)},
\end{align}
where 
\begin{align} \label{G-T}
c_{\rm{eq}}\left( r_{\rm{n}} \right)=&c_{\rm{eq}}\exp \left(\frac{\alpha}{\sqrt[3]{n} } \right);~\alpha=\frac{8\pi r_{0}^{2}\sigma}{3k_{\rm{B}}T }; \nonumber \\
&c_{\rm{eq}}=\exp\left[\left(\mu_{\rm{n}} - \psi \right)/k_{\rm{B}}T\right]
\end{align}
is the thermodynamic equilibrium solute concentration at the interface. Eq.~(\ref{G-T}) is known as the Gibbs-Thomson relation, where $c_{\rm{eq}}$ is a thermodynamic equilibrium solubility, $r_{0}=\sqrt[3]{3\omega_{\rm{n}}/4\pi}$ is an atomic radius and $\alpha$ is a reduced interfacial energy, being the thermodynamic parameter of the model.

Assuming that the solute concentration profile around the cluster remains unperturbed, one can set $c\left(r_{\rm{n}} \right)=\bar{c}$ ($\bar{c}$ being the average bulk solute concentration) in Eq.~(\ref{mu_s}) to obtain from Eq.~(\ref{ratio}) the CNT result $w^{(+)\rm{[CNT]}}_{n, n+1}\Big{/}w^{(-)\rm{[CNT]}}_{n+1, n}=\bar{c}\Big{/}c_{\rm{eq}}\left(r_{\rm{n}} \right)$ (see Eqs.~(\ref{W_a_CNT}), (\ref{W_e_CNT}) below).

The elementary acts of attachment and detachment at the cluster-solution 
interface can be considered as reversible chemical reactions:
\begin{eqnarray} \label{react}
\rm{A}^{\rm{s}} \rightleftharpoons \rm{A}^{\rm{n}},
\end{eqnarray}
where $\rm{A}^{\rm{s}}$ and $\rm{A}^{\rm{n}}$ denote the solute atom in the dissolved and clusterized states, respectively. 
The elementary reactions~(\ref{react}) take place within the spherical interface layer with a radius of $r_{\rm{n}}$ and a thickness equal to the mean elementary jump distance $d$. Let the direct reaction (attachment) be characterized by the reaction rate constant $K_{\rm{a}}$. Since the concentration of dissolved atoms $\rm{A}^{\rm{s}}$ within the interface layer is $c\left(r_{\rm{n}} \right)$, the total rate of attachment within the layer is:
\begin{eqnarray} \label{W_a}
w^{(+)}_{n, n+1}=4\pi r_{\rm{n}}^{2} d K_{\rm{a}} c\left(r_{\rm{n}} \right)/\omega_{\rm{s}},
\end{eqnarray}
where $\omega_{\rm{s}}$ is a mean volume per atom of the solution. Now, using Eq.~(\ref{ratio}), one can derive the rate of detachment as follows:
\begin{eqnarray} \label{W_d}
w^{(-)}_{n+1, n}=4\pi r_{\rm{n}}^{2} d K_{\rm{a}} c_{\rm{eq}}\left(r_{\rm{n}} \right)/\omega_{\rm{s}}.
\end{eqnarray}
The value of the solute concentration at the interface $c\left(r_{\rm{n}} \right)$ can be obtained as a result of solution of the corresponding diffusion boundary problem.

The time-dependent solute concentration profile $c\left(r, t\right)$ around the cluster is subject 
to the next diffusion equation:
\begin{eqnarray} \label{diff_eq}
\partial c/\partial t=-{\rm{div}} j;~j=-D \nabla c,
\end{eqnarray}
where $D$ is a solute diffusion coefficient in the solution.

The normal component~\cite{normal} of the solute flux across the interface is proportional to the difference of the detachment (\ref{W_d}) and attachment (\ref{W_a}) rates: 
\begin{align} \label{flux}
j\left(r_{\rm{n}}, t\right)&=\frac{w^{(-)}_{n+1, n}-w^{(+)}_{n, n+1}}{4\pi r_{\rm{n}}^{2}\omega_{\rm{s}}^{-1}} \nonumber \\
&=\frac{D}{l} \left[c_{\rm{eq}}\left(r_{\rm{n}}\right)-c\left(r_{\rm{n}}, t\right)\right];~l=\frac{D}{d K_{\rm{a}}},
\end{align}
where 
%
%
$l$ is the kinetic parameter of the model with a dimension of length.

One can consider the second boundary condition as follows:
\begin{eqnarray} \label{conc_mean}
c\left( \infty, t \right)=\bar{c}.
\end{eqnarray}
Let the initial solute concentration profile be uniform:
\begin{eqnarray} \label{init}
c\left( r, 0 \right)=\bar{c}.
\end{eqnarray}

The solution of the diffusion equation~(\ref{diff_eq}) with the 
boundary conditions~(\ref{flux}) and 
(\ref{conc_mean}) and the initial condition~(\ref{init}), gives the next expression for the time-dependent solute concentration at the interface (see e.g., Ref.~\onlinecite{Polyanin}):
\begin{align} \label{conc_r_p_t}
&c\left( r_{\rm{n}}, t \right)=\bar{c}  \nonumber \\
&+\frac{r_{\rm{n}}\left[ c_{\rm{eq}}\left( r_{\rm{n}} \right)-\bar{c} \right]}{r_{\rm{n}}+l}\left[ 1- \exp \left( \frac{t}{t_{\rm{diff}}} \right) {\rm{erfc}}\left(\sqrt{\frac{t}{t_{\rm{diff}}}}  \right) \right],
\end{align}
where
\begin{eqnarray} \label{t_diff}
t_{\rm{diff}}=\left( r_{\rm{n}}^{-1}+l^{-1}\right)^{-2}/ D.
\end{eqnarray}
Eq.~(\ref{t_diff}) determines the characteristic timescale of the diffusion process at the interface.

In further considerations of the nucleation process it is convenient to introduce the characteristic nucleation timescale (see Eqs.~(\ref{W_a_small}), (\ref{W_e_small})) and (\ref{W_a_CNT}), (\ref{W_e_CNT}) below):
\begin{eqnarray} \label{t_nucl}
t_{\rm{nucl}}=\left(4\pi r_{0} D c_{\rm{eq}}/\omega_{\rm{s}}\right)^{-1}.
\end{eqnarray}
%

Eq.~(\ref{t_nucl}) determines the characteristic timescale of the nucleation process in both the CNT and this theory. Therefore, the values like $\bar{c}$ vary on this timescale. To ensure the treatment of $\bar{c}$ in the diffusion boundary condition (\ref{conc_mean}) as a constant, the diffusion timescale must be much less than the nucleation one: $t_{\rm{diff}} \ll t_{\rm{nucl}}$. From Eqs.~(\ref{t_diff}) and (\ref{t_nucl}) one finds that this adiabatic approximation holds when the reduced kinetic parameter of the model $\lambda = l\big{/}r_{0}$ is limited as follows:
\begin{eqnarray} \label{adiabatic}
\lambda  \ll \sqrt{\omega_{\rm{s}}/\left(3 \omega_{\rm{n}} c_{\rm{eq}} \right)}.
\end{eqnarray}
In this adiabatic approximation, the solute concentration at the interface approaches its stationary value, which can be obtained from Eq.~(\ref{conc_r_p_t}) in the limit $t \rightarrow \infty$:
\begin{eqnarray} \label{conc_r_p_inf}
c\left( r_{\rm{n}} \right)=\lim_{t \rightarrow \infty} c\left( r_{\rm{n}}, t \right)=\bar{c}+\frac{r_{\rm{n}}\left[ c_{\rm{eq}}\left( r_{\rm{n}} \right)-\bar{c} \right]}{r_{\rm{n}}+l}.
\end{eqnarray}
From Eq.~(\ref{conc_r_p_inf}) one can see that for the finite values of the parameter $l$ the stationary value of the solute concentration at the interface lies between the values of $\bar{c}$ and $c_{\rm{eq}}\left( r_{\rm{n}} \right)$. In the diffusion-limited regime, when the rate of diffusion is much less than the rate of interfacial reactions, $l \ll r_{\rm{n}}$ and $c\left( r_{\rm{n}} \right)=c_{\rm{eq}}\left( r_{\rm{n}} \right)$. In the interface-limited regime, when the rate of interfacial reactions is much less than the rate of diffusion, $l \gg r_{\rm{n}}$ and the CNT assumption $c\left( r_{\rm{n}} \right)=\bar{c}$ holds. 

Below, as an example, we study the process of homogeneous precipitation from a  
supersaturated solution in the framework of the Becker-D\"oring approach~\cite{BD}. 
The distribution function $g \left( n, t \right)$ of the clusters is 
subject to the next kinetic (master) equation~\cite{BD}, valid for $n>1$:
\begin{eqnarray} \label{master_G}
\mathrm{d} g \left( n, t \right)/\mathrm{d} t = J_{n-1, n} - J_{n, n+1}.
\end{eqnarray}
The system of Eqs.~(\ref{master_G}) must be supplemented with 
an additional equation for the reduced bulk solute concentration 
\begin{eqnarray} \label{G1}
g \left( 1, t \right) = \bar{c} \left( t \right) \big{/} c_{\rm{eq}},
\end{eqnarray}
to satisfy the law of 
conservation of the total solute concentration $q$ (see Eq.~(\ref{q}) below):
\begin{eqnarray} \label{master_G1}
\mathrm{d} g \left( 1, t \right) \big{/} \mathrm{d} t = -\sum_{n=2}^{\infty} n \cdot \mathrm{d} g \left( n, t \right) \big{/} \mathrm{d} t.
\end{eqnarray}

From Eqs.~(\ref{W_a}), (\ref{W_d}), taking into account Eqs.~(\ref{conc_r_p_inf}), (\ref{G-T}) and (\ref{t_nucl}), one finds:
\begin{align} \label{W_a_small}
&w^{(+)}_{n, n+1}=\frac{1}{t_{\rm{nucl}}}\frac{\sqrt[3]{n^{2}}}{\lambda} \exp \left( \frac{\alpha}{\sqrt[3]{n}} \right) \times \nonumber \\
&\left\{ 1+\left[ g\left(1, t \right) \exp \left( -\frac{\alpha}{\sqrt[3]{n}} \right)-1\right]\frac{ \lambda}{\lambda+\sqrt[3]{n}} \right\};
\end{align}
\begin{eqnarray} \label{W_e_small}
w^{(-)}_{n+1, n}=\frac{1}{t_{\rm{nucl}}}\frac{\sqrt[3]{n^{2}}}{\lambda} \exp \left( \frac{\alpha}{\sqrt[3]{n}} \right). 
\end{eqnarray}
%

With a different approach the same result has been recently obtained for precipitation from solid solutions~\cite{Borisenko}. Eqs. (\ref{W_a_small}) and (\ref{W_e_small}) for the rates of attachment and detachment at the interface are the central result of the present model. 
They need to be compared to the corresponding equations, derived in the framework of the CNT (see e.g., Ref.~\onlinecite{Slezov} and references therein).
In the present notations, the CNT expressions for the rates of attachment and detachment are as follows:
\begin{eqnarray} \label{W_a_CNT}
w^{(+)\rm{[CNT]}}_{n, n+1}=\frac{1}{t_{\rm{nucl}}}\frac{\sqrt[3]{n^{2}}}{\lambda + \sqrt[3]{n}} g\left(1, t \right);
\end{eqnarray}
\begin{eqnarray} \label{W_e_CNT}
w^{(-)\rm{[CNT]}}_{n+1, n}=\frac{1}{t_{\rm{nucl}}}\frac{\sqrt[3]{n^{2}}}{\lambda + \sqrt[3]{n}} \exp \left( \frac{\alpha}{\sqrt[3]{n}} \right). 
\end{eqnarray}
%
One can note that the pairs of Eqs.~(\ref{W_a_small}), (\ref{W_a_CNT}) and (\ref{W_e_small}), (\ref{W_e_CNT}) are asymptotically equivalent for $\lambda \gg \sqrt[3]{n}$. Great values of the parameter $\lambda$ correspond to the interface-limited precipitation regime, when the mobility of dissolved atoms at the interface is small compared to the bulk one. In this case the solute concentration profile around the cluster is flat and one can get from Eq.~(\ref{conc_r_p_inf}) $c\left( r_{\rm{n}} \right)=\bar{c}$, in agreement with the CNT assumption. Therefore, the present theory contains the CNT in itself as a limiting case. At the same time, the value $\mathrm{d} n\big{/}\mathrm{d} t = w^{(+)}_{n, n+1}-w^{(-)}_{n+1, n}$, which is determined by the diffusion-controlled net solute flux in the solution, is equal in both the CNT and this theory. That is why both theories give the same results in the asymptotic coarsening regime, but differ in the range of ultrafine clusters (see Figs.~\ref{Fig:3} and \ref{Fig:4} below).

Under the condition of detailed balance, when the flux of clusters in 
the dimension space (\ref{J}) turns to zero for any 
$n$ (see Eq.~(\ref{J_0})), the equilibrium distribution function~(\ref{Boltzmann_distr}) can be expressed as follows:
\begin{eqnarray}\label{g_0}
g_{\rm{eq}}\left(n\right)=
\begin{cases}
g_{\rm{eq}}\left(1\right),~n=1; \\ 
g_{\rm{eq}}\left(1\right) \prod \limits_{i=2}^{n} w^{(+)}_{i-1, i}\big{/}w^{(-)}_{i, i-1},~n \geq 2.
\end{cases}
\end{eqnarray}

In the range of undersaturated and saturated solute concentrations, i.e. for $0 \le g_{\rm{eq}}\left(1\right) \le 1$, the equilibrium distribution function is finite: $\lim_{n \to \infty}g_{\rm{eq}}\left(n\right)=0$, while in the range of supersaturated concentrations, i.e. for $g_{\rm{eq}}\left(1\right) > 1$, the equilibrium distribution function is divergent: $\lim_{n \to \infty}g_{\rm{eq}}\left(n\right)=\infty$. 

A total solute concentration (expressed in the units 
of $c_{\rm{eq}}$) can be calculated as 
follows:
\begin{eqnarray} \label{q}
q = \sum_{n=1}^{\infty} n \cdot g \left( n, t \right).
\end{eqnarray}

In the limiting case $g_{\rm{eq}}^{*}\left(1\right) =1$, corresponding to the saturated solute concentration, Eq.~(\ref{q}) 
with $g \left( n, t \right)=g_{\rm{eq}}^{*}\left(n\right)$ can be utilized to calculate the total solubility, taking into 
account both the single dissolved atoms and thermodynamic heterophase fluctuations (ultrafine clusters):
\begin{eqnarray} \label{q*}
q^{*} = \sum_{n=1}^{\infty} n \cdot g_{\rm{eq}}^{*}\left(n\right).
\end{eqnarray}
\begin{figure}
\centerline{\includegraphics[width=0.9\textwidth]{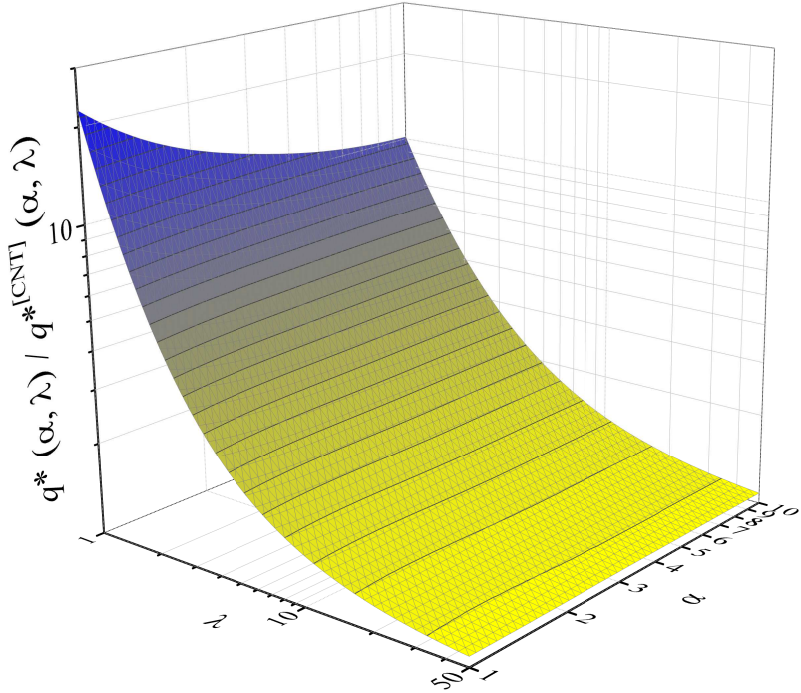}}
\caption{\label{Fig:1} Relative total solubility $q^{*}/q^{*\rm{[CNT]}}$ vs. model parameters $\alpha$ and $\lambda$.}
\end{figure}

Fig.~\ref{Fig:1} shows the relative total solubility $q^{*}/q^{*\rm{[CNT]}}$ as a function of two dimensionless model parameters $\alpha$ and $\lambda$. The values corresponding to the present theory and the CNT are calculated from Eq.~(\ref{q*}) with the attachment and detachment rates given by Eqs.~(\ref{W_a_small}), (\ref{W_e_small}) and (\ref{W_a_CNT}), (\ref{W_e_CNT}), respectively. From Fig.~\ref{Fig:1} one can see that the CNT systematically underestimates the total solubility and that this discrepancy vanishes with 
an increase of $\lambda$. 

In the steady-state nucleation regime the flux of clusters in 
the dimension space~(\ref{J}) is constant for any 
$n$: 
\begin{eqnarray} \label{J_const}
J_{n, n+1}=J=const,~\forall n.
\end{eqnarray}

The steady-state distribution function in this case is as follows:
\begin{eqnarray} \label{g_J}
\frac{g_{J}\left(n\right)}{g_{\rm{eq}}\left(n\right)}=\frac{g_{J}\left(1\right)}{g_{\rm{eq}}\left(1\right)}-J\sum \limits_{i=1}^{n} \left[w^{(+)}_{i, i+1}g_{\rm{eq}}\left(i\right)\right]^{-1}.
\end{eqnarray}
Following Zeldovich~\cite{Zeldovich}, one can set $g_{J}\left(1\right)\big{/}g_{\rm{eq}}\left(1\right)=1$ and $\lim_{n \to \infty}g_{J}\left(n\right)\big{/}g_{\rm{eq}}\left(n\right)=0$ to obtain from Eq.~(\ref{g_J}) an explicit form for the steady-state flux of clusters in the dimension space (the nucleation rate):
\begin{eqnarray}\label{J_nucl}
J=\left\{\sum \limits_{i=1}^{\infty} \left[w^{(+)}_{i, i+1}g_{\rm{eq}}\left(i\right)\right]^{-1}\right\}^{-1}.
\end{eqnarray}
\begin{figure}
\centerline{\includegraphics[width=0.9\textwidth]{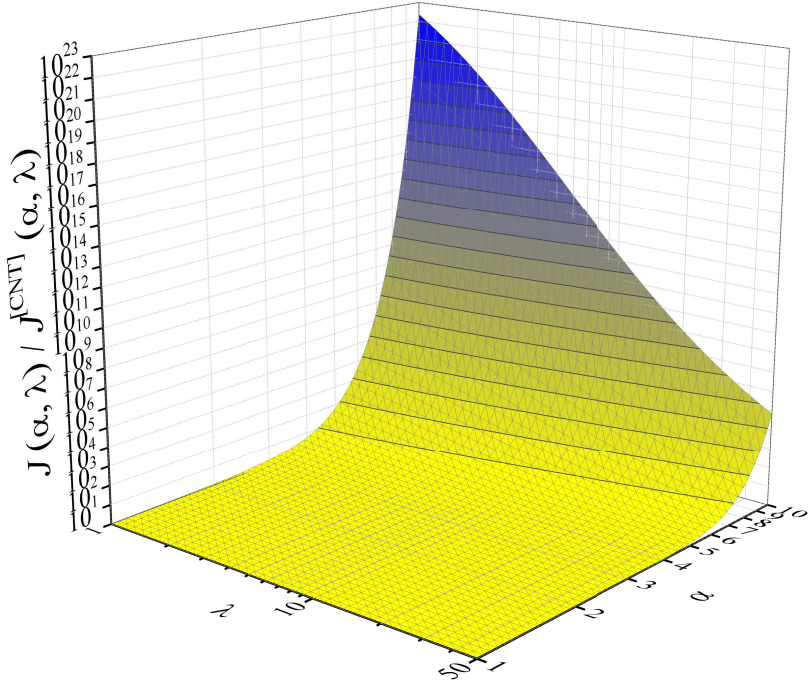}}
\caption{\label{Fig:2} Relative nucleation rate $J/J^{\rm{[CNT]}}$ at $g_{\rm{eq}}\left(1\right)=15$ vs. model parameters $\alpha$ and $\lambda$.}
\end{figure}

Fig.~\ref{Fig:2} shows the relative nucleation rate $J/J^{\rm{[CNT]}}$ at the degree of supersaturation $g_{\rm{eq}}\left(1\right)=15$ as a function of two dimensionless model parameters $\alpha$ and $\lambda$. The values corresponding to the present theory and the CNT are calculated from Eq.~(\ref{J_nucl}) with the attachment and detachment rates given by Eqs.~(\ref{W_a_small}), (\ref{W_e_small}) and (\ref{W_a_CNT}), (\ref{W_e_CNT}), respectively. From Fig.~\ref{Fig:2} one can see that the CNT systematically underestimates the nucleation rate. While in the range of small values of $\alpha$ this discrepancy is negligibly small, it quickly increases and may achieve many orders of magnitude with an increase of $\alpha$ and a decrease of $\lambda$.

Below we compare the results of the present model for precipitation kinetics with those of the CNT, for the same values of the total solute concentration~(\ref{q}) $q=10^{4}$ and the model parameters $\alpha=3$ and $\lambda=1$.  
In each calculation, the homogeneous state of a solution (only single atoms, no clusters) is taken as an initial condition.
\begin{figure}
\centerline{\includegraphics[width=0.9\textwidth]{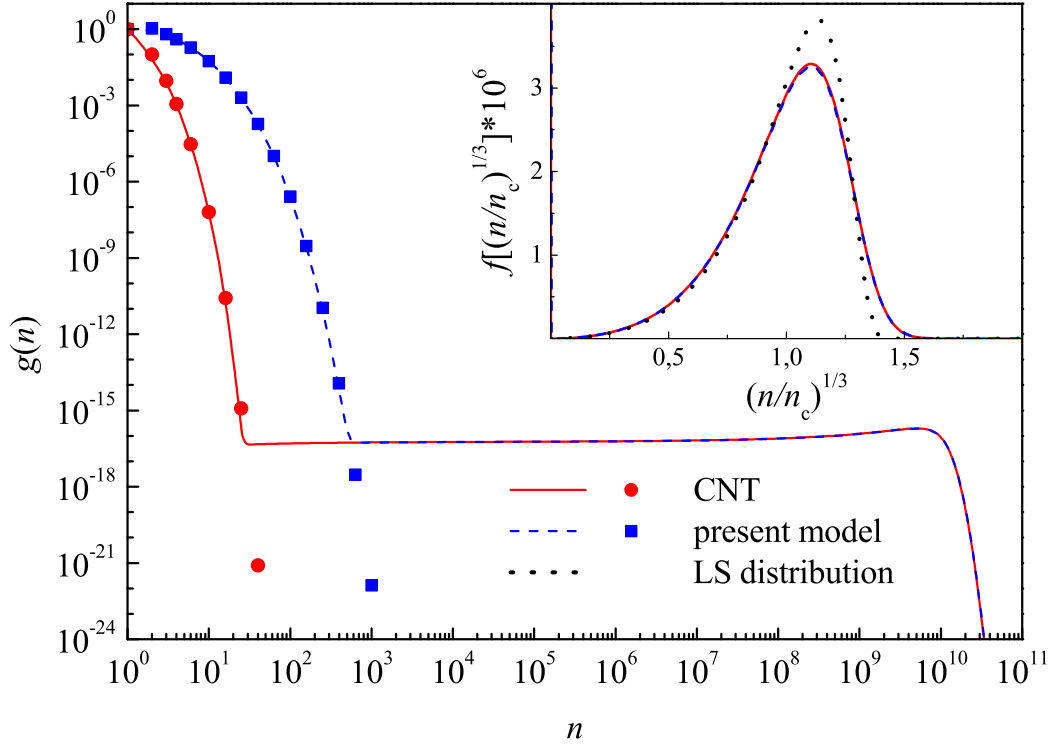}}
\caption{\label{Fig:3} A distribution of clusters at $t/t_{\rm{nucl}}=10^{10}$, calculated in the framework of the CNT and this model. Symbols show the corresponding equilibrium distributions of heterophase fluctuations. The inset shows the calculated distributions in the reduced coordinates together with the LS distribution function.}
\end{figure}
The dashed and solid curves in Fig.~\ref{Fig:3} show the solutions of the system of Eqs.~(\ref{master_G}), (\ref{master_G1}) at $t/t_{\rm{nucl}}=10^{10}$, with the rates of attachment and detachment, given by this model (Eqs.~(\ref{W_a_small}),  (\ref{W_e_small})) and the CNT (Eqs.~(\ref{W_a_CNT}), (\ref{W_e_CNT})), respectively. The low-$n$ steep parts of the curves describe the thermodynamic heterophase fluctuations and coincide with the corresponding equilibrium distributions~(\ref{g_0}), shown by symbols. At the same time, the high-$n$ flat parts describe the clusters, which evolve according to the Lifshitz-Slyozov-Wagner (LSW) theory~\cite{LS, Wagner}. The inset in Fig.~\ref{Fig:3} shows the numerical data recalculated according to the rule $f\left( x \right) \, \mathrm{d} x = g \left( n \right) \, \mathrm{d} n$, where $x=\sqrt[3]{n/n_{\rm{c}}}$ is the reduced cluster size and $n_{\rm{c}}=\left(\alpha/\ln g\left(1, t \right)\right)^3$ is the cluster critical size. One can see that the calculated curves approach the Lifshitz-Slyozov (LS) distribution function~\cite{LS}.
One can see that this model gives a much wider range of heterophase fluctuations than the CNT does. This result qualitatively explains why this model gives larger values for the total solubility (see Fig.~\ref{Fig:1}) and the nucleation rate (see Fig.~\ref{Fig:2}), compared to the CNT results. At the same time, in the high-$n$ 
range both models give identical results, in agreement with the LSW theory. This happens because the difference of the attachment and detachment rates (the net solute flux, which solely enters the LSW theory) is equal in both this theory and the CNT. 
%
\begin{figure}
\centerline{\includegraphics[width=0.9\textwidth]{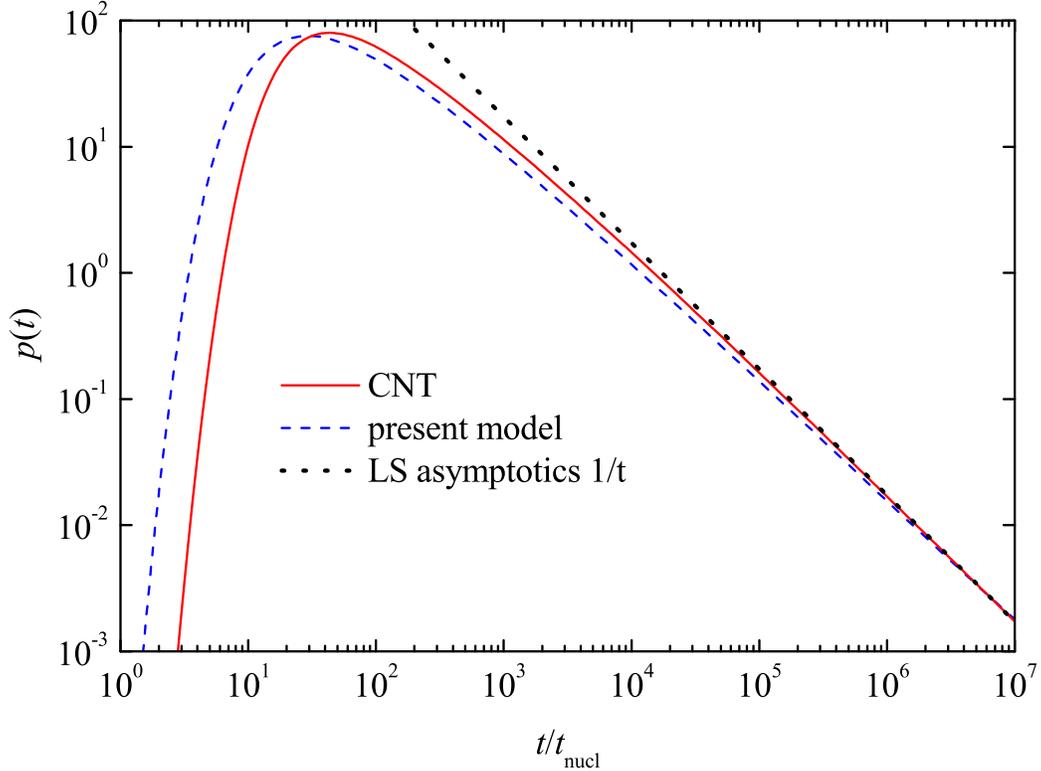}}
\caption{\label{Fig:4} A concentration of clusters as a function of time (\ref{p}), calculated in the framework of the CNT and this model together with the LS asymptotic law $1/t$.}
\end{figure}

Fig.~\ref{Fig:4} shows the "observable" concentration of clusters in the range $n \geq n_{\rm{min}}$: 
\begin{eqnarray} \label{p}
p \left( t \right) = \sum_{n=n_{\rm{min}}}^{\infty}  g \left( n, t \right),
\end{eqnarray}
where $n_{\rm{min}}$ is a lower limit cutoff, practically set by the resolution limit of an observation instrument. In Fig.~\ref{Fig:4} $n_{\rm{min}}=60$.
One can see that within the present model the nucleation stage starts earlier than within the CNT, and at the coarsening stage the asymptotic LS power law $p(t) \propto 1/t$ is achieved later within the present model than within the CNT.

It is worth noting that the assumption $n \gg 1$ made during derivation of Eq.~(\ref{ratio}) makes the present approach (just like the CNT one) inapplicable in practice for small clusters with $n \gtrsim 1$. Instead, one has to obtain the first several points of the distribution function $g_{\rm{eq}}\left(n\right)$ in the range $1 \leq n \leq n^{*}$ from experimental or simulation data and then, using the recursive structure of Eq.~(\ref{g_0}), proceed to higher $n$. This "hybrid" distribution function then reads:
\begin{eqnarray}\label{g_hybrid}
g_{\rm{eq}}\left(n\right)=
\begin{cases}
g_{\rm{eq}}^{\rm{exp}}\left(n\right),~1 \leq n \leq n^{*}; \\ 
g_{\rm{eq}}^{\rm{exp}}\left(n^{*}\right) \prod \limits_{i=n^{*}+1}^{n} w^{(+)}_{i-1, i}\big{/}w^{(-)}_{i, i-1},~n > n^{*}.
\end{cases}
\end{eqnarray}

In summary, the proposed model of precipitation from supersaturated solutions is grounded on the basic assumptions of the CNT, with an exception of the postulate of a homogeneous distribution of solute atoms around the clusters. Taking into account the diffusion processes around the clusters allows the correct derivation of the attachment and detachment rates at the interface, which enter the master kinetic equation and govern the process of precipitation. Numerical calculations of the total solubility and the nucleation rate show that the CNT systematically underestimates both of these values and that the corresponding discrepancies, depending on the values of the model parameters, may achieve many orders of magnitude. However, the discrepancies vanish in the limit of the interface-limited precipitation, when the solute concentration at the interface approaches its value in the bulk.

The author is grateful to Dr. A. Turkin for sharing his numerical code and for helpful discussions.


\begin{thebibliography}{99}
\bibitem{normal} The normal unit vector at the interface is assumed to be directed outside of the cluster.
\end{thebibliography}
\end{document}